\documentclass[10pt,a4paper]{article}
\usepackage{amsmath}
\usepackage{amssymb}
\usepackage{graphicx}

\def\aa{A\&A }

\def\aj{AJ }

\usepackage{bm}
\usepackage{color}
\usepackage{pgfplots}
\usepackage{hyperref}
\begin{document}

\setcounter{figure}{0}
\setcounter{table}{0}
\setcounter{footnote}{0}
\setcounter{equation}{0}

\vspace*{0.5cm}

\noindent {\Large RANGE, DOPPLER AND ASTROMETRIC  OBSERVABLES COMPUTED FROM TIME TRANSFER FUNCTIONS: A SURVEY}
\vspace*{0.7cm}

\noindent\hspace*{1.5cm} A. HEES$^1$, S. BERTONE$^{2,3}$, C. LE PONCIN-LAFITTE$^2$, P. TEYSSANDIER$^2$\\
\noindent\hspace*{1.5cm} $^1$ Department of Mathematics, Rhodes University, Grahamstown 6140, South Africa\\
\noindent\hspace*{1.5cm} e-mail: A.Hees@ru.ac.za\\
\noindent\hspace*{1.5cm} $^2$  Observatoire de Paris, SYRTE, CNRS/UMR 8630, LNE, UPMC\\
\noindent\hspace*{1.5cm}  61 avenue de l'Observatoire, F-75014 Paris, France\\
\noindent\hspace*{1.5cm} e-mail: \{stefano.bertone,christophe.leponcin,pierre.teyssandier\}@obspm.fr\\
\noindent\hspace*{1.5cm} $^3$  currently at Astronomical Institute, University of Bern\\

\vspace*{0.5cm}

\noindent {\large ABSTRACT.} Determining range, Doppler and astrometric observables is of crucial interest for modelling and analyzing space observations. We recall how these observables can be computed when the travel time of a light ray is known as a function of the positions of the emitter and the receiver for a  given instant of reception (or emission). For a long time, such a function--called a reception (or emission) time transfer function--has been almost exclusively calculated by integrating the null geodesic equations describing the light rays. However, other methods avoiding such an integration have been considerably developped in the last twelve years. We give a survey of the analytical results obtained with these new methods up to the third order in the gravitational constant $G$ for a mass monopole. We briefly discuss the case of quasi-conjunctions, where higher-order enhanced terms must be taken into account for correctly calculating the effects. We summarize the results obtained at the first order in $G$ when the multipole structure and the motion of an axisymmetric body is taken into account. We present some applications to on-going or future missions like Gaia and Juno. We give a short review of the recent works devoted to the numerical estimates of the time transfer functions and their derivatives.
\vspace*{1cm}

\noindent {\large 1. OBSERVABLES COMPUTABLE FROM TIME TRANSFER FUNCTIONS}

\smallskip

Many observations in the Solar System rest on the measurement of the travel time of light rays. Modelling the light propagation requires a mathematical tool defined as follows. Assume that space-time is covered by a single system of coordinates $x^0=ct, \bm x=(x^i)$, where $i=1,2,3$. Consider a light ray emitted at time $t_{\scriptscriptstyle A}$ at a point of spatial coordinates ${\bm x}_{\scriptscriptstyle A}$ and received at time $t_{\scriptscriptstyle B}$ at a point of spatial coordinates ${\bm x}_{\scriptscriptstyle B}$. Here, light rays are null geodesic paths (light propagating in a vacuum). The light travel time $t_{\scriptscriptstyle B}-t_{\scriptscriptstyle A}$ may be regarded as a function of the variables $\bm x_{\scriptscriptstyle A},t_{\scriptscriptstyle B}, \bm x_{\scriptscriptstyle B}$, so that one can write
\begin{equation} \label{T}
t_{\scriptscriptstyle B}-t_{\scriptscriptstyle A} ={\cal T}_{r}(\bm x_{\scriptscriptstyle A},t_{\scriptscriptstyle B}, \bm x_{\scriptscriptstyle B}).
\end{equation}

${\cal T}_{r}$ may be called the ``(reception) time transfer function" (TTF)\footnote{In this communication, we generally omit the term ``reception" for the sake of brevity. Note that similar results can be derived from the ``(emission) time transfer function" $\mathcal T_e$ defined by $t_{\scriptscriptstyle B}-t_{\scriptscriptstyle A} ={\cal T}_{e}(t_{\scriptscriptstyle A}, \bm x_{\scriptscriptstyle A}, \bm x_{\scriptscriptstyle B})$.}. As we shall see below, the interest of this function is not confined to the range experiments: knowing ${\cal T}_{r}$ is sufficient for modelling observations based on the Doppler-tracking or the gravitational bending of light (astrometry).

1. Suppose that the above-mentioned signal is exchanged between two observers ${\cal O}_{\scriptscriptstyle A}$ and  ${\cal O}_{\scriptscriptstyle B}$. 
Let $\nu_{\scriptscriptstyle A}$ and $\nu_{\scriptscriptstyle B}$ be the frequencies of the signal as measured at $(ct_{\scriptscriptstyle A} , \bm x_{\scriptscriptstyle A})$ by ${\cal O}_{\scriptscriptstyle A}$ and at $(ct_{\scriptscriptstyle B} , \bm x_{\scriptscriptstyle B})$ by ${\cal O}_{\scriptscriptstyle B}$, respectively. The ratio $\nu_{\scriptscriptstyle B}/\nu_{\scriptscriptstyle A}$ is given by (see, e.g., Teyssandier et al 2008b and Refs. therein)
\begin{equation} \label{fAB2}
\frac{\nu_{\scriptscriptstyle B}}{\nu_{\scriptscriptstyle A}}=\frac{[(g_{00}+2g_{0i}\beta^{i}+g_{ij}\beta^{i}\beta^{j})^{1/2}]_{x_{\scriptscriptstyle A}}}{[(g_{00}+2g_{0i}\beta^{i}+g_{ij}\beta^{i}\beta^{j})^{1/2}]_{x_{\scriptscriptstyle B}}} \,\frac{(k_{0})_{x_{\scriptscriptstyle B}}}{(k_{0})_{x_{\scriptscriptstyle A}}}\, \frac{1+(\beta^{i} \widehat{k}_{i})_{x_{\scriptscriptstyle B}}}{1+(\beta^{i} \widehat{k}_{i})_{x_{\scriptscriptstyle A}}},
\end{equation}
where the quantities $g_{\alpha\beta}$ are the components of the metric, $\beta^{i}_{x_{\scriptscriptstyle A}}=[dx^i_{\scriptscriptstyle A}/cdt]_{t_{\scriptscriptstyle A}}$ and $\beta^{i}_{x_{\scriptscriptstyle B}}=[dx^i_{\scriptscriptstyle B}/cdt]_{t_{\scriptscriptstyle B}}$ are the coordinate velocities divided by $c$ of ${\cal O}_{\scriptscriptstyle A}$ at time $t_{\scriptscriptstyle A}$ and ${\cal O}_{\scriptscriptstyle B}$ at time $t_{\scriptscriptstyle B}$, respectively. The quantities $\widehat k_i$ are defined by $\widehat{k}_{i}={k}_{i}/{k}_{0}$, where the $k_{\alpha}$ are the covariant components of the vector $k^{\mu}$ tangent to the light ray described by affine parametric equations. 
One has (see Le Poncin-Lafitte et al 2004)
\begin{equation}\label{eq:hatk}
	\left(\widehat k_i\right)_{\scriptscriptstyle A}=c\frac{\partial \mathcal T_r}{\partial x^i_{\scriptscriptstyle A}}, \qquad 	\left(\widehat k_i\right)_{\scriptscriptstyle B}=-c\frac{\partial \mathcal T_r}{\partial x^i_{\scriptscriptstyle B}}\left[1-\frac{\partial \mathcal T_r}{\partial t_{\scriptscriptstyle B}}\right]^{-1}, \qquad \frac{\left(k_0\right)_{\scriptscriptstyle B}}{\left(k_0\right)_{\scriptscriptstyle A}}=1-\frac{\partial \mathcal T_r}{\partial t_{\scriptscriptstyle B}}.
\end{equation}
Substituting these relations in (\ref{fAB2}) yields $\nu_{\scriptscriptstyle B}/\nu_{\scriptscriptstyle A}$ in terms of the derivatives of the TTF as follows
\begin{equation} \label{fABT}
\frac{\nu_{\scriptscriptstyle B}}{\nu_{\scriptscriptstyle A}}=\frac{[(g_{00}+2g_{0i}\beta^{i}+g_{ij}\beta^{i}\beta^{j})^{1/2}]_{x_{\scriptscriptstyle A}}}{[(g_{00}+2g_{0i}\beta^{i}+g_{ij}\beta^{i}\beta^{j})^{1/2}]_{x_{\scriptscriptstyle B}}} \, \frac{1-\frac{\partial {\cal T}_{r}}{\partial t_{\scriptscriptstyle B}}-c\beta^{i}_{x_{\scriptscriptstyle B}}\frac{\partial {\cal T}_{r}}{\partial x^{i}_{\scriptscriptstyle B}} }{1+c\beta^{i}_{x_{\scriptscriptstyle A}}\frac{\partial {\cal T}_{r}}{\partial x^{i}_{\scriptscriptstyle A}}},
\end{equation}
a formula which can also be inferred without using (\ref{fAB2}), as it is shown in Hees et al 2012.

2. Let $\{\lambda_{\underline{\alpha}}, \underline{\alpha}=0, 1, 2, 3\}$ be an orthonormal comoving tetrad attached to ${\cal O}_{\scriptscriptstyle B}$ ($\lambda_{\underline{0}}$ coincides with the unit 4-velocity vector of ${\cal O}_{\scriptscriptstyle B}$). The direction of the light ray as measured by ${\cal O}_{\scriptscriptstyle B}$ is defined by a unit vector proportional to the orthogonal projection of $k^{\mu}$ on the rest frame of ${\cal O}_{\scriptscriptstyle B}$ at $x_{\scriptscriptstyle B}$. The spatial components $n^{\underline{i}}$ of this vector in the basis $\{\lambda_{\underline{i}}\}$ 
is given by (see, e.g., Brumberg 1991)
\begin{equation} \label{ni}
n^{\underline{i}}=-\left(\frac{\lambda_{\underline{i}}^{0}+\lambda_{\underline{i}}^{j}\,\widehat{k}_{j}}{\lambda_{\underline{0}}^{0}+\lambda_{\underline{0}}^{j}\,\widehat{k}_{j}}\right)_{x_{\scriptscriptstyle B}},
\end{equation}
where $\lambda_{\underline{\alpha}}^{\mu}$ denote the components of the 4-vector $\lambda_{\underline{\alpha}}$ in the natural basis associated to the coordinates ($x^{\mu}$). It follows from (\ref{eq:hatk}) that each $n^{\underline{i}}$ can be expressed in terms of the derivatives of the TTF. 

An analogous conclusion can be drawn for the angular separation $\phi$ between two point-like sources $S$ and $S'$ as measured by ${\cal O}_{\scriptscriptstyle B}$ at $x_{\scriptscriptstyle B}$. Indeed, one has (Teyssandier \& Le Poncin-Lafitte 2006)
\begin{equation}\label{eq:angSep}
	\sin^2\frac{\phi}{2}=-\frac{1}{4}\left[\frac{\left(g_{00}+2g_{0k}\beta^k+g_{kl}\beta^k\beta^l\right)g^{ij}(\widehat k_i -\widehat k_i')(\widehat k_j -\widehat k_j')}{(1+\beta^m \widehat k_m)(1+\beta^l \widehat k_l')}\right]_{\scriptscriptstyle B} \; ,
\end{equation}
where the quantities $\widehat{k}_{i}$ and $\widehat{k}'_{i}$ are related to the light rays arriving from $S$ and $S'$, respectively.

\vspace*{0.7cm}

\noindent {\large 2. A SURVEY OF THE METHODS PROPOSED FOR CALCULATING THE TTFs}

\smallskip 

Two approaches exist to determine the light propagation in metric theories of gravity. The most widespread method consists in solving the null geodesic equations. Analytical solutions have been developed within the first post-Newtonian (1pN) or first post-Minkowksian (1pM) approximation dealing with static monopoles (Shapiro 1964), static mass multipole moments (Kopeikin 1997), moving monopoles (Kopeikin \& Sch\"affer 1999 and Klioner 2003a), moving multipole moments (Kopeikin \& Makarov 2007),... After the pioneering papers by Richter \& Matzner 1983 and Brumberg 1987, an analytical solution has been derived within the 2pM approximation for a static monopole, with a metric containing three arbitrary post-Newtonian parameters (Klioner \& Zschocke 2010). Finally, the gravitational deflection of the image of a star when observed at a finite distance from a static monopole has been obtained up to the 2pM order in Ashby \& Bertotti 2010. On the other hand, a numerical treatment based on a shooting method has been proposed in San Miguel 2007.

The other approach enables to determine the TTFs without integrating the null geodesic equations. Initially grounded on Synge's world function (see John 1975 for the Schwarzschild space-time, and then Linet \& Teyssandier 2002, Le Poncin-Lafitte et al 2004 for much more general cases), this approach is now based on the direct determination of the TTFs (Teyssandier \& Le Poncin-Lafitte 2008a).

\vspace*{0.7cm}
\noindent {\large 3. POST-MINKOSWKIAN EXPANSION OF THE TTF} 
\smallskip

We assume that the metric may be expanded in a series in powers of the gravitational constant $G$:
\begin{equation} \label{gmnG}
g_{\mu\nu}(x,G)=\eta_{\mu\nu}+\sum_{n=1}^{\infty} g_{\mu\nu}^{(n)}(x,G),
\end{equation}
where $\eta_{\mu\nu}=$ diag $\{1, -1, -1, -1\}$ is the Minkowski metric and $g_{\mu\nu}^{(n)}(x, G)$ stands for the term of order $G^n$. Then, it may be supposed that ${\cal T}_{r}$ is represented by an asymptotic expansion in a series in powers of $G$:
\begin{equation} \label{expT}
{\cal T}_{r}(\bm x_{\scriptscriptstyle A},t_{\scriptscriptstyle B}, \bm x_{\scriptscriptstyle B})=\frac{R_{\scriptscriptstyle AB}}{c}+\sum_{n=1}^{\infty}{\cal T}_{r}^{(n)}(\bm x_{\scriptscriptstyle A},t_{\scriptscriptstyle B}, \bm x_{\scriptscriptstyle B}),
\end{equation}          
where $R_{\scriptscriptstyle AB}=\vert\bm x_{\scriptscriptstyle B}-\bm x_{\scriptscriptstyle A}\vert$ and ${\cal T}_{r}^{(n)}$ stands for the perturbation term of order $G^n$. It is shown in Teyssandier \& Le Poncin-Lafitte 2008a that each ${\cal T}_{r}^{(n)}$ can be expressed by an iterative procedure as a line integral whose the integrand involves only the terms $g_{\mu\nu}^{(k)}$ and ${\cal T}_{r}^{(l)}$ such that $k\leq n-1$, $l\leq n-1$, with an integration taken along the straight line passing through $x_{\scriptscriptstyle B}$ defined by
\begin{equation} \label{zl}
x^{\alpha}=z^{\alpha}(\lambda), \quad z^{0}(\lambda)=x^{0}_{\scriptscriptstyle B}-\lambda R_{\scriptscriptstyle AB},\quad z^{i}(\lambda)=x_{\scriptscriptstyle B}^{i}-
\lambda (x_{\scriptscriptstyle B}^{i}-x_{\scriptscriptstyle A}^{i}), \quad 0\leq \lambda\leq 1.
\end{equation} 
So, computing the TTFs never requires the knowledge of the real null geodesics followed by the photons. 

\vspace*{0.7cm}

\noindent {\large 4. APPLICATION TO STATIC, SPHERICALLY SYMMETRIC SPACE-TIMES}

\smallskip

The procedure outlined in section 3 allows the determination of the TTF and the direction of light propagation in a static spherically symmetric space-time at any order in $G$ (Teyssandier 2014). This determination can also be obtained by an iterative solution of an integro-differential equation derived from the null geodesic equations (Linet \& Teyssandier 2013). Denoting by $M$ the mass of the central body and assuming the metric to be a generalization of the Schwarzschild $ds^2$ written in the form  
\begin{equation} \label{ds2}
ds^2=\left(1 - \frac{2m}{r} + 2\beta \frac{m^2}{r^2} -\frac{3}{2}\beta_{3} \frac{m^3}{r^3}+\cdots\right)(dx^0)^2-\left(1 + 2 \gamma \frac{m}{r} +\frac{3}{2} \epsilon\frac{m^2}{r^2}+\frac{1}{2}\gamma_{3} \frac{m^3}{r^3}+\cdots\right)d\bm x^2,
\end{equation}         
where $r=\vert \bm x\vert$, $m=GM/c^2$ and the coefficients $\beta , \beta_3, \gamma , \epsilon , \gamma_3,$ are post-Newtonian parameters equal to 1 in general relativity, the two methods lead to expressions\footnote{Note that owing to the static character of the metric, ${\cal T}_{r}$ does not depend on $t_{\scriptscriptstyle B}$. So we may remove the index $r$.} as follow for the first three terms in Eq.~(\ref{expT}):
\begin{subequations}
\begin{eqnarray} 
& &\!\!{\cal T}^{(1)}( \bm x_{\scriptscriptstyle A}, \bm x_{\scriptscriptstyle B}) = \frac{(1+\gamma)m}{c}\ln\left(\frac{r_{\scriptscriptstyle A}+r_{\scriptscriptstyle B}+R_{\scriptscriptstyle AB}}{r_{\scriptscriptstyle A}+r_{\scriptscriptstyle B}-R_{\scriptscriptstyle AB}}\right),\qquad \label{T1} \\ 
& & \nonumber \\
& &\!\!{\cal T}^{(2)}( \bm x_{\scriptscriptstyle A}, \bm x_{\scriptscriptstyle B}) = \frac{m^2}{r_{\scriptscriptstyle A}r_{\scriptscriptstyle B}} \frac{R_{\scriptscriptstyle AB}}{c}\bigg[ \kappa\frac{\arccos \bm n_{\scriptscriptstyle A}.\bm n_{\scriptscriptstyle B}}{\vert\bm n_{\scriptscriptstyle A}\times\bm n_{\scriptscriptstyle B}\vert}-
\frac{(1+\gamma)^2}{1+\bm n_{\scriptscriptstyle A}.\bm n_{\scriptscriptstyle B}}\bigg],\qquad\label{T2}\\ 
& & \nonumber \\
& &\!\!{\cal T}^{(3)}(\bm x_{\scriptscriptstyle A},\bm x_{\scriptscriptstyle B})=\frac{m^3}{r_{\scriptscriptstyle A}r_{\scriptscriptstyle B}}\left( \frac{1}{r_{\scriptscriptstyle A}}+\frac{1}{r_{\scriptscriptstyle B}}\right)\frac{R_{\scriptscriptstyle AB}}{c(1+\bm n_{\scriptscriptstyle A}.\bm n_{\scriptscriptstyle B})} \bigg\lbrack \kappa_3 -(1+\gamma)\kappa\frac{\arccos \bm n_{\scriptscriptstyle A} .\bm n_{\scriptscriptstyle B}}{\vert\bm n_{\scriptscriptstyle A}\times\bm n_{\scriptscriptstyle B}\vert}+\frac{(1+\gamma )^3}{1+\bm n_{\scriptscriptstyle A}.\bm n_{\scriptscriptstyle B}}\bigg\rbrack,\qquad \label{T3} 
\end{eqnarray}
\end{subequations} 
where $\bm n_{\scriptscriptstyle A}=\bm x_{\scriptscriptstyle A}/r_{\scriptscriptstyle A}$, $\bm n_{\scriptscriptstyle B}=\bm x_{\scriptscriptstyle B}/r_{\scriptscriptstyle B}$ and $\kappa=2(1+\gamma)-\beta+3/4\varepsilon$, $\kappa_3=2\kappa-2\beta(1+\gamma)+(3\beta_3 + \gamma_3)/4$.

Equation (\ref{T1}) is equivalent to the well-kown formula due to Shapiro and (\ref{T2}) recovers the expression already obtained in Teyssandier \& Le Poncin-Lafitte 2008a, and then in Klioner \& Zschocke 2010. On the other hand, (\ref{T3}) is a recent result and shows the fecondity of the new procedures. 

It follows from Eqs.~(\ref{T1})-(\ref{T3}) that at least for $n\leq 3$, an enhancement of the contribution proportional to $(1+\gamma)^n$ appears in configurations of quasi-conjunction, i.e. when the unit 3-vectors $\bm n_{\scriptscriptstyle A}$ and $\bm n_{\scriptscriptstyle B}$ are almost opposite ($1+\bm n_{\scriptscriptstyle A}.\bm n_{\scriptscriptstyle B}\sim 0$). A result inferred in Ashby \& Bertotti 2010 by an `asymptotic reasoning' is thus rigorously confirmed. The 2pM enhanced term in~(\ref{T2}) will be required for analyzing data in future missions like for example BepiColombo (Iess et al 2009), as it may be seen on Figs. 2 and 3 in Hees et al 2014a. The 3pM enhanced contribution from the Sun may reach 30 ps for a light ray grazing the Sun (see Table 1 in Linet \& Teyssandier 2013). Taking this contribution into account will therefore be necessary for modelling space mission proposals like ODYSSEY (Christophe 2009), LATOR (Turyshev 2009) or ASTROD (Braxmaier et al 2012), designed to measure the 1pN parameter $\gamma$ at the level of $10^{-7}$-$10^{-8}$.

The light deflection has been calculated and discussed within the 2pM approximation in Klioner \& Zschocke 2010, Ashby \& Bertotti 2010 and Teyssandier 2012. The enhanced 2pM term, proportional to $(1+\gamma)^2$, can reach 16 microarcsecond ($\mu$as) for a ray grazing Jupiter (see right of Fig. 2) and is therefore required in the analysis of Gaia data (see, e.g., de Bruijne 2012). In Teyssandier \& Linet 2013 and Hees et al 2014a, it is noted that  for a ray grazing the Sun, the 2pM and 3pM enhanced contributions amount to 3 milliarcsecond (mas) and 12 $\mu$as, respectively. The last value is to be compared with the 2pM contribution due to the 2pN parameter $\kappa$, as illustrated on the left of Fig. 2.

\begin{figure}[hbt]
\begin{center}
\includegraphics[width=0.45\textwidth]{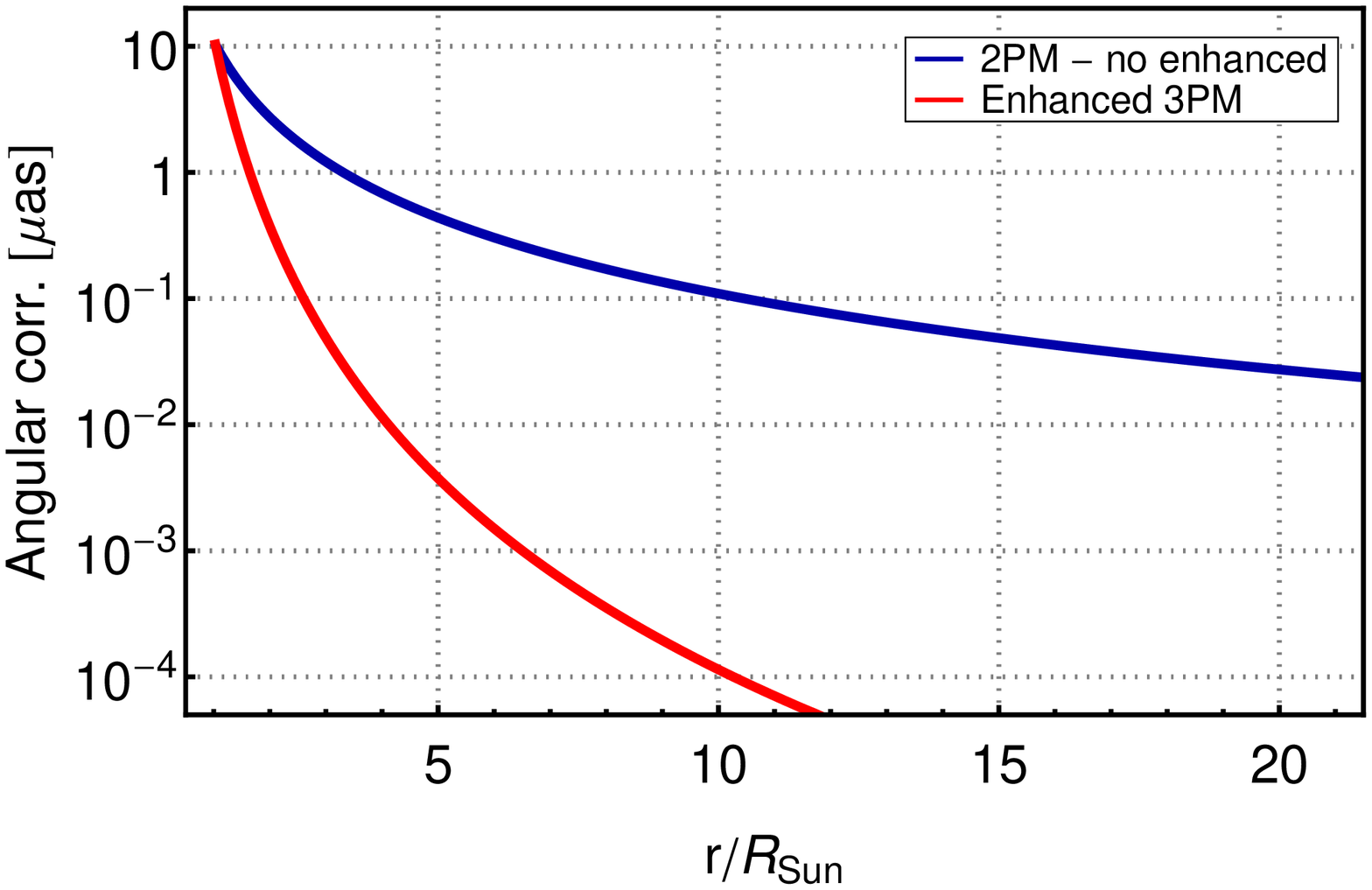}\hfill
\includegraphics[width=0.45\textwidth]{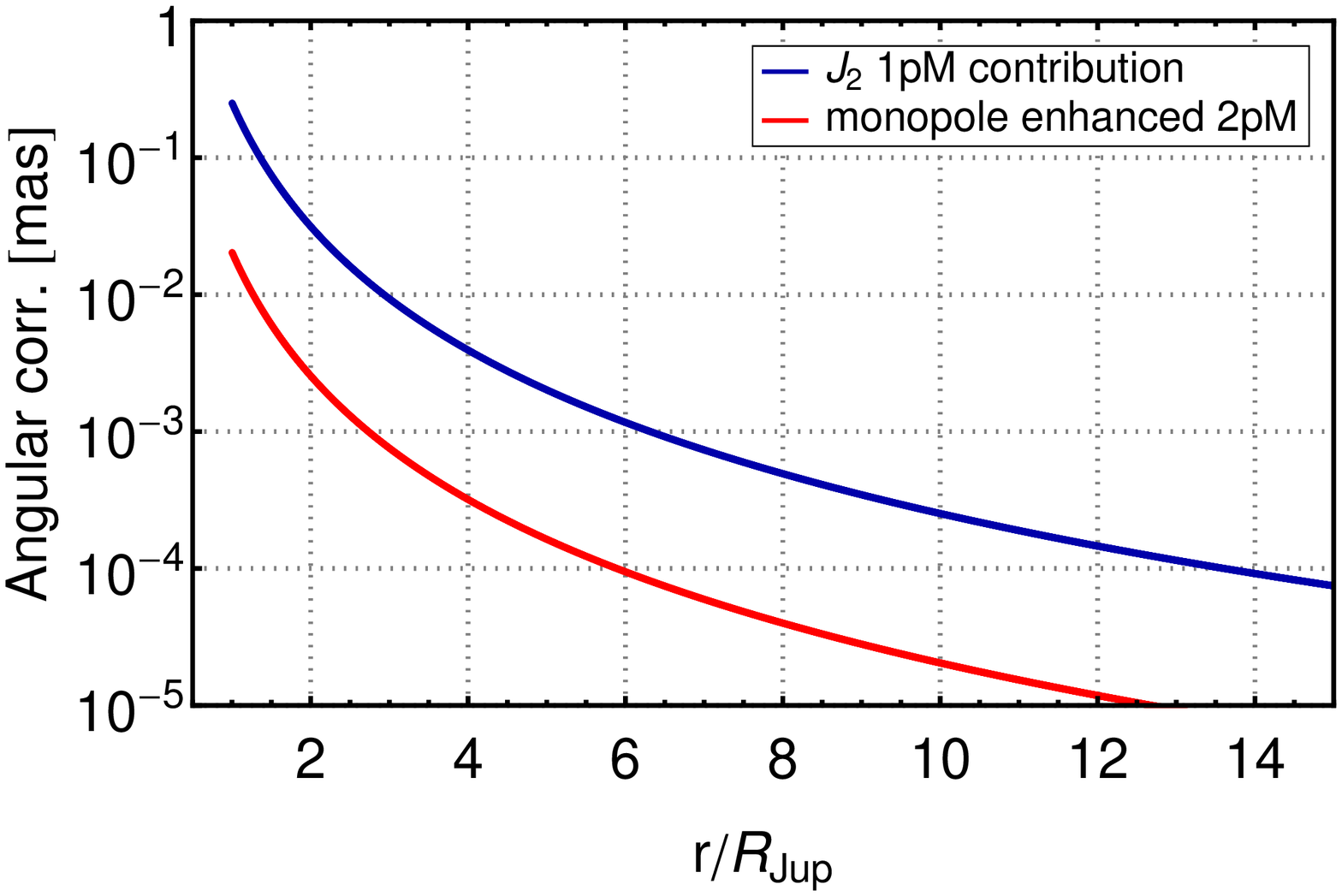}
\caption{Left: Contribution of the 2pM term proportional to $\kappa$ and the 3pM enhanced term on the light deflection for a Sun grazing  ray. -- Right: Contribution of Jupiter $J_2$ at 1pM order and contribution of the enhanced 2pM Jupiter monopole term on the  deflection of a Jupiter grazing light ray.}
\label{fig:deflection}
\end{center}
\end{figure}

\vspace*{0.7cm}

\noindent {\large 5. EFFECTS DUE TO THE ASPHERICITY AND/OR THE MOTION OF BODIES}

\smallskip

\begin{figure}[hbt]
\begin{center}
\includegraphics[width=0.45\textwidth]{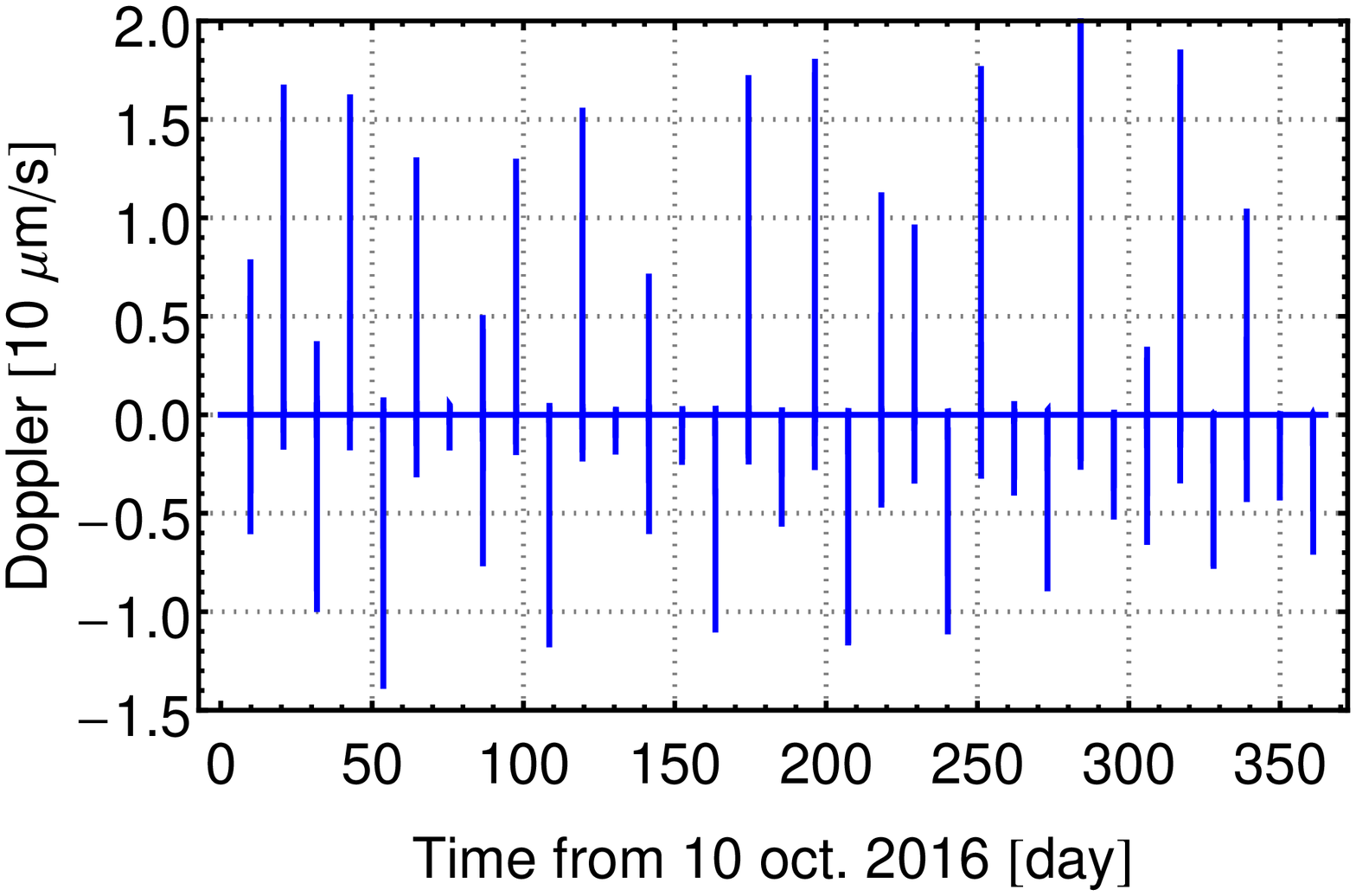}\hfill
\includegraphics[width=0.45\textwidth]{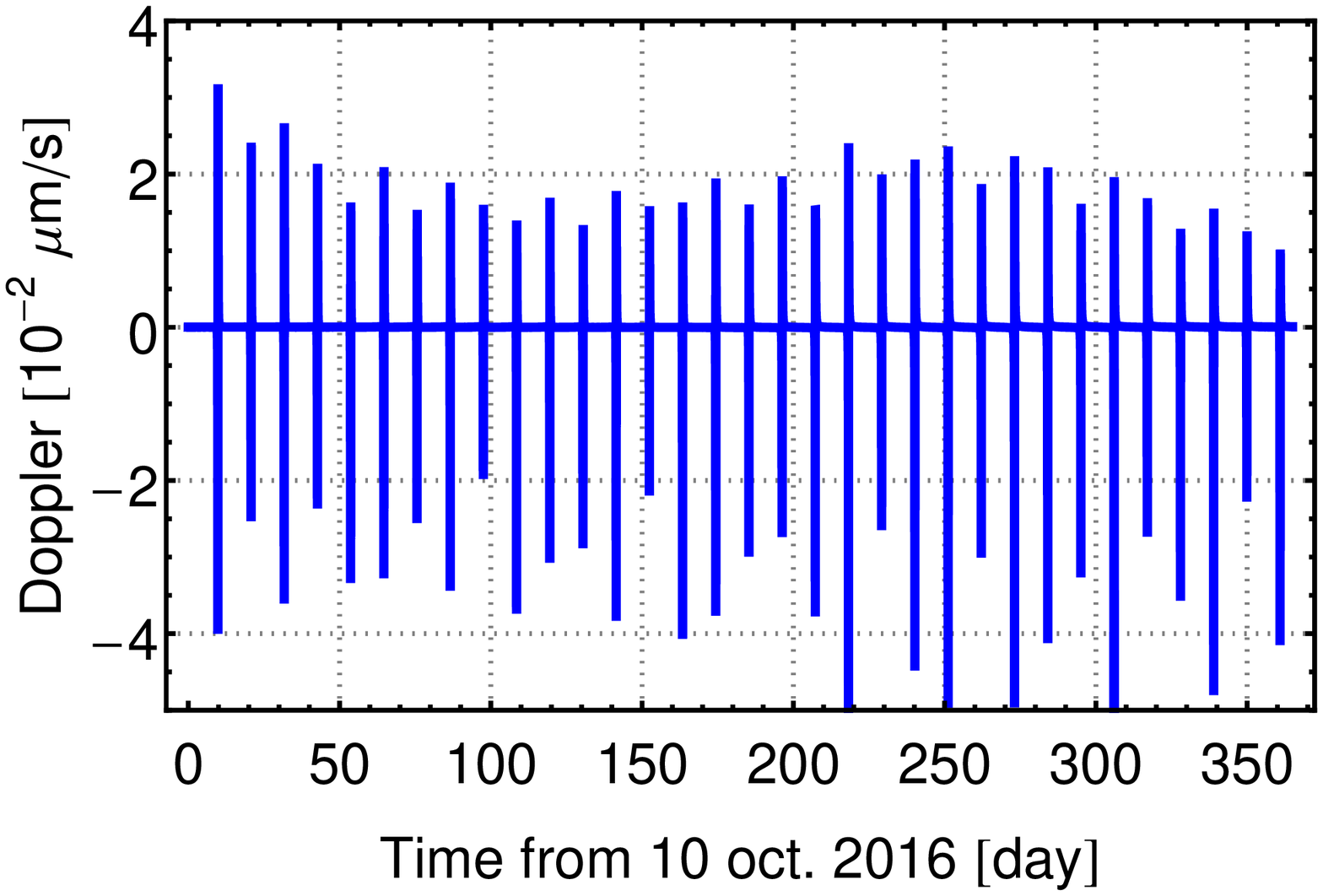}
\caption{Left: Effect of Jupiter $J_2$ on a Doppler link between Juno and Earth. Right: Effect of Jupiter's velocity on a Doppler link between Juno and Earth.}
\label{fig:juno}
\end{center}
\end{figure}

The gravitational potential of an axisymmetric body is parametrized amongst others by its mass multipole moments $J_n$. Using a property previously  established in Kopeikin 1997 and recovered later (see Teyssandier et al 2008b and Refs. therein), explicit formulas for the contributions of each $J_n$ to the TTF and its first derivatives have been given in Le Poncin-Lafitte \& Teyssandier 2008. Thus, it becomes possible to calculate the influence of any $J_n$ on the gravitational light deflection. These results generalize the expressions previously obtained in various papers for $n=1$ and $n=2$ (see, e.g., Klioner 2003b, Kopeikin \& Makarov 2007, and Refs. therein). Recall that the  Jupiter $J_2$ must be taken into account in the analysis of Gaia (see Crosta \& Mignard 2006 and Refs. therein) or VLBI observations (see the right of Fig.~\ref{fig:deflection}) since it produces a deflection amounting to 240 $\mu$as for a grazing light ray. A similar conclusion holds for the Juno mission (see Anderson et al 2004) since it is shown in Hees et al 2014b that the influence of the quadrupole moment of Jupiter reaches the level of the cm for the range and the level of 10 $\mu$m/s for the Doppler (see left of Fig.~\ref{fig:juno}). Some of these effects will be relevant in the data reduction since the expected accuracies for Juno are of 10 cm on the range and 1 $\mu$m/s on the Doppler.

The procedure outlined in section~3 noticeably facilitates the determination of the TTF of a uniformly moving axisymmetric body within the 1pM approximation. Denote by $\tilde {\mathcal T}^{(1)}_r$ the 1pM TTF corresponding to the body at rest. When this body is uniformly moving with a coordinate velocity $\bm v =c\bm \beta$, it is shown in Hees et al 2014b that the 1pM TTF can be written as
\begin{equation} \label{T1mb}
 \mathcal T_r^{(1)}(\bm x_{\scriptscriptstyle A},t_{\scriptscriptstyle B},\bm x_{\scriptscriptstyle B})=\Gamma(1-\bm N_{AB}.\bm \beta) \tilde  {\mathcal T}_r^{(1)}(\bm R_{\scriptscriptstyle A}+\Gamma R_{AB} \bm \beta,\bm R_{\scriptscriptstyle B})\; ,
\end{equation}
where $\Gamma=(1-\beta^2)^{-1/2}$ is the Lorentz factor and
\begin{equation}
 \bm R_X=\bm x_X-\bm x_p(t_0)+\frac{\Gamma^2}{1+\Gamma}\bm \beta\left[\bm\beta.(\bm x_X-\bm x_p(t_0))\right] -\Gamma \bm v (t_{\scriptscriptstyle B}-t_0)\; ,
\end{equation}
with $\bm x_p(t_0)$ being the position of the deflecting body at an arbitrary time $t_0$ usually chosen between $t_{\scriptscriptstyle B}-R_{AB}/c$ and $t_{\scriptscriptstyle B}$. This recent and general result is particularly simple. The first derivatives of the right-hand side of (\ref{T1mb}) are easily calculated. For a moving monopole, using Eq.~(\ref{T1}) for $\tilde  {\mathcal T}_r^{(1)}$ and Eq.~(\ref{T1mb}) gives
\begin{equation}
 \mathcal T^{(1)}_{r}(\bm x_{\scriptscriptstyle A},t_{\scriptscriptstyle B},\bm x_{\scriptscriptstyle B})=(1+\gamma) m \Gamma (1-\bm\beta.\bm N_{AB})\ln \frac{\left|\bm R_{\scriptscriptstyle A} + \Gamma R_{AB} \bm \beta\right|+R_{\scriptscriptstyle B}+\Gamma R_{AB}(1-\bm \beta.\bm N_{AB})}{\left|\bm R_{\scriptscriptstyle A} + \Gamma R_{AB} \bm \beta\right|+R_{\scriptscriptstyle B}-\Gamma R_{AB}(1-\bm \beta.\bm N_{AB})}.
\end{equation}
This formula recovers the expression obtained in Kopeikin \& Sch\"affer 1999 and Klioner 2003a using longer calculations. A low velocity expansion of this result is obtained in Bertone et al 2014. To finish, let us mention that using a similar method but a symmetric trace free (STF) decomposition of the gravitational potential, Soffel \& Han 2014 have also determined the expression of the TTF produced by a moving body with arbitrary static multipoles, but their result is only valid in the slow velocity approximation. 

In Hees et al 2014b, these results are applied in the context of the Juno mission to discuss the effects of the mass and the quadrupole moment of Jupiter when the motion of this planet is taken into account. The effect of the motion of Jupiter's monopole is represented on the right of Fig.~\ref{fig:juno}. This contribution is smaller than the expected Juno Doppler accuracy and can safely be ignored in the reduction of the observations. Nevertheless, it is important to point out that this numerical estimate depends highly on the geometry of the probe orbit and should be reassessed in the context of other space missions. In particular, this contribution depends on the quantity $\bm \beta.\bm N_{AB}$ and on the presence of conjunctions (which is not the case for Juno owing its polar orbit, but will be the case in other missions). The deflection of light produced by the motion of Jupiter monopole is of the order of 0.04 $\mu$as for a grazing light ray and can safely be ignored for current observations.

\vspace*{0.7cm}

\noindent {\large 6. NUMERICAL DETERMINATION OF THE TTFs AND THEIR DERIVATIVES}

\smallskip

The TTF formalism lends itself well to the numerical simulations of the light propagation in curved space-time. This is useful when no analytical expressions can be found or when systematic comparisons  of the propagation of light in different space-times are discussed. This approach is fully developed within the 2pM approximation in Hees et al 2014a. The iterative procedure mentioned in Sect.~3 gives 
\begin{subequations}
\begin{eqnarray}
&&\mathcal T_r^{(1)}=\int_0^1 n\left[z^\alpha(\lambda);g^{(1)}_{\alpha\beta},R_{AB}\right]d\lambda,\label{T1a}\\ 
&& \frac{\partial  \mathcal T_r^{(1)}}{\partial x^i_{A/B}}=\int_0^1 n_{A/B}\left[z^\alpha(\lambda);g^{(1)}_{\alpha\beta},g^{(1)}_{\alpha\beta,\sigma},\bm x_{A},\bm x_{B}\right]d\lambda \label{dT1},
\end{eqnarray}
\end{subequations}
for $\mathcal T_r^{(1)}$ and its first derivatives, and then
\begin{subequations}
\begin{eqnarray}
&&  \mathcal T_r^{(2)}=\int_0^1 \int_0^1 l\left[z^\alpha(\mu\lambda);g^{(2)}_{\alpha\beta},g^{(1)}_{\alpha\beta},g^{(1)}_{\alpha\beta,\sigma},\bm x_{A},\bm x_{B}\right]d\mu \, d\lambda \, ,\\
&& \frac{\partial  \mathcal T_r^{(2)}}{\partial x^i_{A/B}}=\int_0^1\int_0^1 l_{A/B}\left[z^\alpha(\mu\lambda);g^{(2)}_{\alpha\beta},g^{(2)}_{\alpha\beta,\sigma},g^{(1)}_{\alpha\beta},g^{(1)}_{\alpha\beta,\sigma},g^{(1)}_{\alpha\beta,\sigma\delta},\bm x_{A},\bm x_{B}\right]d\mu\, d\lambda\; ,
\end{eqnarray}
\end{subequations}
for $\mathcal T_r^{(2)}$ and its first derivatives, where the functions $n$, $n_{\scriptscriptstyle A}$, $n_{\scriptscriptstyle B}$, $l$, $l_{\scriptscriptstyle A}$ and $l_{\scriptscriptstyle B}$ can be explicitly written (see Hees et al 2014a). All the integrations are taken over the straight line defined by Eqs.~(\ref{zl}). 

This kind of procedure avoids the numerical integration of the full set of geodesic equations, which is unnecessarily time consuming since we are only concerned by a single `time function'. It has been successfully applied to simulate range, Doppler and astrometric observations within some alternative theories of gravity in order to find signatures differing from the predictions of general relativity (Hees et al 2012 and Hees et al 2014c), and more recently to compute the propagation of light in the field of arbitrarily moving monopoles, when no analytical solution is available (Hees et al 2014b). 

\vspace*{0.7cm}

\noindent {\large 7. CONCLUSION}

\smallskip

This survey shows that the TTF formalism is a powerful tool for computing the range, Doppler and astrometric (VLBI) observables involved in  Solar System experiments. The iterative method summarized in section 3 is very effective in deriving analytical and numerical solutions. The simplicity of this method relies mainly on the fact that one never has to determine the real trajectory of the photon in order to perform an explicit calculation of the TTF. We have reviewed some of the analytical expressions derived using this formalism. This method has been successfully applied to determine the light propagation in a static spherically symmetric space-time up to the 3pM order and a generic procedure enabling to compute higher order terms has been developed. It has also been applied to determine the influence of the motion and asphericity of bodies on the light propagation. The result is obtained by simple calculations. We have assessed the influence of different terms in the observation of space missions like Gaia or Juno. Finally, the TTF formalism turns out to be very well adapted to the numerical simulations of the effects observable in the Solar System.

\vspace*{0.7cm}
\textit{Acknowledgements.} A.H. thanks the organizers for financial support to attend this meeting. The authors are grateful for the financial support of CNRS/GRAM and Observatoire de Paris/GPHYS.
\vspace*{0.7cm}

\noindent {\large 8. REFERENCES}

{

\leftskip=5mm
\parindent=-5mm

\smallskip

Anderson, J. D., Lau, E. L., Schubert, G., Palguta, J. L., 2004, Bull. Am. Astron. Soc. 36, 1094.

Ashby, N., Bertotti, B., 2010, Class. and Quantum Grav. 27, 145013.

Bertone S., et al, 2014, Class. and Quantum Grav. 31, 015021.

Braxmaier, C., et al, 2012, Exp. Astron. 34, 181.

Brumberg, V. A., 1987 Kinematics Phys. Celest. Bodies, 3, pp. 6-12.

Brumberg, C., 1991, ``Essential relativistic celestial mechanics", Adam Hilger.

Christophe, B., et al, 2009, Exp. Astron. 23, 529.

Crosta, M. T., Mignard, F., 2006, Class. Quantum Grav. 23, 4853.

de Bruijne, J. H. J., 2012, Science performance of Gaia, ESA's space astrometry mission, arXiv:1201.3238.

Hees, A., Lamine, B., Reynaud, S., et al 2012, Class. and Quantum Grav. 29, 235027.

Hees, A., Bertone, S., Le Poncin-Lafitte, C., 2014a, Phys. Rev. D 89, 064045.

Hees, A., Bertone, S., Le Poncin-Lafitte, C., 2014b, Phys. Rev. D 90, 084020.

Hees, A., Folkner, W., Jacobson, R., Park, R., 2014c, Phys. Rev. D 89, 102002.

Iess, L., Asmar, S., Tortora, P., 2009, Acta Astronaut. 65, 666.

John, R. W., 1975, Exp. Tech. Phys., 23, pp. 127-140.

Klioner, S., 2003a, \aa 404, 783.

Klioner, S., 2003b, \aj 125, 1580.

Klioner, S., Zschocke, S., 2010, Class. and Quantum Grav. 27, 075015. 

Kopeikin, S., 1997, J. of Math. Physics 38, 2587.

Kopeikin, S., Sch\"affer G., 1999, Phys. Rev. D 60, 124002.

Kopeikin, S., Makarov, V., 2007, Phys. Rev. D 75, 062002.

Le Poncin-Lafitte, C., Linet, B., Teyssandier P., 2004, Class. and Quantum Grav. 21, 4463.

Le Poncin-Lafitte, C., Teyssandier P., 2008, Phys. Rev. D 77, 044029.

Linet, B., Teyssandier, P., 2002, Phys. Rev. D 66, 024045.

Linet, B., Teyssandier, P., 2013, Class. and Quantum Grav. 30, 175008.

Richter, G., Matzner, R., 1983, Phys. Rev. D 28, 3007.

San Miguel, A., 2007, Gen. Rel. and Grav. 39, 2025.

Shapiro, I., 1964, Phys. Rev. Letters 13, 789.

Soffel, M., Han, W.-B., 2014, arXiv:1409.3743.


Teyssandier, P., Le Poncin-Lafitte, C., 2006, arXiv:gr-qc/0611078.

Teyssandier, P., Le Poncin-Lafitte, C., 2008a, Class. and Quantum Grav. 25, 145020.

Teyssandier, P., Le Poncin-Lafitte, C., Linet, B., 2008b, in Lasers, Clocks and Drag-Free Control: 

\hspace{5mm}Exploration of Relativistic Gravity in Space, p. 153, Springer.

Teyssandier P., 2012, Class. and Quantum Grav. 29, 245010.

Teyssandier, P., Linet B., 2014, Proc. Journ\'ees Syst\`emes de r\'ef\'erence 2013, Paris; arXiv:1312.3510.

Teyssandier P., 2014, in Frontiers in Relativistic Celestial Mechanics, S.M. Kopeikin ed., vol. 2, p.1, De Grutyer, Berlin; arXiv:1407.4361.

Turyshev, S. G., et al, 2009, Exp. Astron. 27, 27.

}

\end{document}